%% file: main.tex
\documentclass[runningheads]{llncs}

\usepackage{graphicx}
\usepackage{booktabs}
\usepackage{siunitx}
\usepackage{multirow}
\usepackage{hyperref}
\begin{document}
\title{On the Influence of Reading Sequences on Knowledge Gain during Web Search}
\titlerunning{Reading Sequences in SAL}
\author{Wolfgang Gritz\inst{1}\orcidID{0000-0003-1668-3304} \and
Anett Hoppe\inst{1,2}\orcidID{0000-0002-1452-9509} \and
Ralph Ewerth\inst{1,2}\orcidID{0000-0003-0918-6297}}

\institute{TIB -- Leibniz Information Centre for Science and Technology, Welfengarten 1B, Hannover, 30167, Germany \and 
Leibniz University Hannover, L3S Research Center, Appelstraße 9A, Hannover, 30167, Germany
\email{wolfgang.gritz@tib.eu}
}

\authorrunning{W. Gritz et al.}
\maketitle              % typeset the header of the contribution
\begin{abstract}
Nowadays, learning increasingly involves the usage of search engines and web resources.
The related interdisciplinary research field search as learning aims to understand how people learn on the web. 
Previous work has investigated several feature classes to predict, for instance, the expected knowledge gain during web search. 
Therein, eye-tracking features have not been extensively studied so far. 
In this paper, we extend a previously used reading model from a line-based one to one that can detect reading sequences across multiple lines. 
We use publicly available study data from a web-based learning task to examine the relationship between our feature set and the participants' test scores. 
Our findings demonstrate that learners with higher knowledge gain spent significantly more time reading, and processing more words in total. 
We also find evidence that faster reading at the expense of more backward regressions may be an indicator of better web-based learning.
We make our code publicly available at \url{https://github.com/TIBHannover/reading_web_search/}.

\end{abstract}

\section{Introduction}
The research field \textit{Search as Learning} (SAL) examines knowledge acquisition processes during and with the help of web search.
Whereas conventional information retrieval systems focus on satisfying an information need, the objective here is to determine how individuals can be best supported to perform an engaging and successful learning session.
The learning success is often measured by \textit{knowledge gain}, that is, the change in the learner's knowledge state achieved through the search~\cite{Hoppe2018,Machado2020}.

Past SAL research on knowledge gain has followed various directions, e.g., the time frames in which learning happens~\cite{Roy20Exploring,Zein2023Evolution}, the characteristics of a web resource indicating suitability for learning~\cite{Yu2018,Yu2021,Otto2021,Gritz2021,tang2021domain}, and the \textit{user behavior}.
The research to \textit{user behavior} is based on e.g., the analysis of input queries~\cite{Collins-Thompson2016}, navigation logs~\cite{Eickhoff2014lessons}, or other behavioral features~\cite{Gadiraju2018,Yu2018,Yu2021,tang2021domain}.
One branch of user behavior attempts to associate reading behavior with learning using eye-tracking data.
Bhattacharya and Gwizdka~\cite{Bhattacharya2019} conducted a web search study to compare the reading behavior of more and less successful learners.
For this, they relied on the line-based reading model from previous work~\cite{Cole2011,Cole2013}.

In this paper, we suggest to extend their reading model in two regards:
(1) to verify that the fixations actually apply to words on the web page and (2) to also cover line breaks and paragraph breaks.
We derive a set of novel features based on textual content pages to capture reading behavior.
For the evaluation, we use the publicly available \emph{SaL-Lightning dataset}~\cite{lightning}, which consists of data from an explorative web search study.
We aim to answer the following questions:
(1) Can we reproduce the results for content pages in Bhattacharya and Gwizdka~\cite{Bhattacharya2019} on other study data?
(2) Can we model a learner's prior knowledge by behavioral reading features?
(3) Can we use behavioral reading features to determine learning success?
Contrary to Bhattacharya and Gwizdka~\cite{Bhattacharya2019}, our experimental results show that learners, who spent more time reading, processed more words, and exhibited faster reading with more backward regressions, achieved higher knowledge gain and a better web-based learning performance.

The remainder of the paper is structured as follows: In Section~\ref{sec:rw}, we summarize related work that investigates indicators for successful web search.
In Section~\ref{sec:method}, we describe how reading sequences are identified from individual fixations of the eye-tracking data.
In Section~\ref{sec:eval}, we analyze the mean differences between more and less successful learners.
Finally, we summarize the results in Section~\ref{sec:conclusions} and discuss the limitations of the approach.

\section{Related Work}\label{sec:rw}
Central research questions in SAL concern the description and detection of typical SAL-related user behavior and the design of optimized retrieval and ranking algorithms for learning~\cite{RULK,RULKNE,Rokicki2022Learning}.
For determining characteristics that can indicate learning success, previous work has examined both user behavior and visited resources (i.e., web pages, their appearance, and content).
For example, Vakkari~\cite{Vakkari16} conducted a survey that captured features indicating potential knowledge gain during the search process.
In the context of resource-centric features, Otto et al.~\cite{Otto2021} examined the number and type of graphics on websites.
In contrast, Gritz et al.~\cite{Gritz2021} studied the influence of the textual complexity of the page.
In this context, Pardi et al.~\cite{Pardi2020} found evidence that text-based web pages seem to have a more substantial influence on a user's knowledge gain.
Yu et al.~\cite{Yu2018,Yu2021} and Tang et al.~\cite{tang2021domain} also used textual and HTML features in order to predict knowledge gain.
But besides resource features, they additionally suggest behavioral features of the user.
Similar to Eickhoff et al.~\cite{Eickhoff2014lessons}, they also explore the entered queries, as well as the click and scroll behavior.
Other approaches aim to predict the perceived webpage relevance~\cite{Bhattacharya2020a,Bhattacharya2020b} for the user, or even of single paragraphs~\cite{Barz2021}.
Cole et al.~\cite{Cole2011,Cole2013} and Bhattacharya and Gwizdka~\cite{Bhattacharya2019} determined the influence of reading sequences on learning by analyzing eye-tracking data.
Therefore, they used a line-based reading model to determine, i.e., temporally successive fixations that are associated with a reading process of the user.
It is based on the assumption that reading sequences can be determined by successive fixations in a roughly horizontal trajectory.
However, the reading model ignores the underlying text and treats sequences as single lines.

\section{Detection of Reading Sequences}\label{sec:method}
In this section, we present our method for the detection of reading sequences that can deal with line breaks, reading sequences across several lines, and paragraph breaks. 
Our method is based on the reading model of Cole et al.~\cite{Cole2011} which has been used several times in other studies~\cite{Bhattacharya2019,Cole2011,Cole2013}.
The model relies only on the coordinates of fixations as input without considering the actual web page.
We extend the approach by incorporating data from the rendered web pages provided by the \emph{reading protocol}~\cite{ReadingProtocol}.
It receives a web page and resolution as input and then renders the web page.
As an output, we get the position of the individual words as individual bounding boxes with a continuous index $I_{w}$ that also crosses paragraph boundaries.
This enables not only the filtering of falsely detected reading sequences (i.e., fixations not on textual content) but also the recognition of reading sequences over multiple lines.

\paragraph{(Para)foveal region definition:}
Crucial for the detection of reading sequences are the values for the foveal and parafoveal regions on the screen plane.
The foveal region is responsible for clear central focus during reading, while the parafoveal region detects surrounding words.
In the algorithm, they determine the maximum distance between two fixations in order to be considered as a reading sequence.
Since the values for the foveal and parafoveal region radius $r_{foveal}$ and $r_{parafoveal}$ are not contained in the used dataset~\cite{lightning}, we calculate them in a preprocessing step.
We use the following values from the literature~\cite{Gwizdka19}: $d_{foveal}{=}2^\circ$, $d_{parafoveal}{=}7^\circ$, and $d_{display}{=}65cm$, i.e., the diameter of the foveal or parafoveal region on the screen and the average distance from participant to display.
The dataset description also gives a screen diagonal of $24"$ and $1920{\times}1080$ resolution.
This yields the values $r_{foveal}{\approx}41$ and $r_{parafoveal}{\approx}185$ in pixels.

\paragraph{Definitions:}
Let: $F_{t}$ denote the fixation at time $t$;
$W$ represent the set of words in the text;
$L$ be a reading line;
$d(F_{t}, w)$ be the Euclidean distance between the fixation $F_{t}$ and word $w\in W$;
$PR(F_{t})$ be the parafoveal region for a fixation $F_{t}$.
We define the parafoveal region to the right as $r_{parafoveal}$ and to the left, top, and bottom as $r_{foveal}$.

\paragraph{Candidate selection:}
Eye-tracking data are noisy and may have errors due to measurement inaccuracies and aggregation of measurement points to fixations.
Therefore, we do not necessarily assign the word with the smallest distance to a fixation, but determine a set of the most likely candidates.
We consider only those words that fall within a certain radius $r$ around the fixation point:
    $C(F_{t}) = \{w \,|\, d(F_{t}, w) \leq r\}$.
We assume a maximum error of one complete foveal region in each direction, resulting in $r=2\cdot r_{foveal}$.
We sort the candidate words in ascending order based on their distances to the fixation.
If no candidates exist, we assume the fixation was not on text and cannot be part of a \emph{reading sequence}.

\paragraph{Sequential Line Processing}
We first check, independent of the words, whether successive fixations are approximately in the same line.
For each fixation $F_{t}$, we check if the next fixation $F_{t+1}$ falls within the parafoveal region $PR(F_{t})$:
(1) If $F_{t+1}$ is within $PR(F_{t})$, we add $F_{t+1}$ to the current reading line $L$, or
(2) if $F_{t+1}$ is not within $PR(F_{t})$, we conclude that the current reading line $L$ has ended.
For each reading line $L$, we apply the Viterbi algorithm to find the most
likely
words for the fixations.
The cost function is defined as:
\begin{equation}
    \mathcal{L} = C_{R}(F_{t})^2 + (1 - C_{I}(F_{t+1}) - C_{I}(F_{t}))^2
\end{equation}
where $C_{R}(F_{t})$ represents the rank of the candidate words $C(F_{t})$ for the fixation $F_{t}$, and $C_{I}(F_{t+1})$ the index of the candidate words on the web page.
The two summands are intended to represent a tradeoff between (1) the spacing of fixations and words and (2) the spacing of words in the text.

\paragraph{Reading Sequence Definition:}
After the assignment of words to fixations, we only consider the indices to define reading sequences.
We define that a fixation $F_{t}$ belongs to a \emph{reading sequence} $RS$, if:
(1) $C_{I}(F_{t-1}) \leq C_{I}(F_{t}) \leq C_{I}(F_{t-1}) + 4$, indicating that the next fixation is at most four words away, or
(2) $\min(RS) \leq C_{I}(F_{t}) \leq \max(RS)$, indicating that the fixation lies within the previously read \emph{reading sequence}.
We refer to the latter case as \emph{regression}.
This approach takes into account fixations outside the text and imperfect accuracy of eye-tracking data, including misassignment of fixations to nearby words on different lines.

\section{Experimental Results}\label{sec:eval}
\paragraph{Dataset:}
For the evaluation, we rely on the publicly available \emph{SaL-Lightning dataset}~\cite{lightning} of an exploratory web search with $N{=}106$ participants.
Since the study language was German, the reading direction was from left to right.
The participants were supposed to gather knowledge about the formation of thunder and lightning.
In addition to the tracking data, the authors also provided the original web page data.
Since we primarily aim to investigate the influence of reading sequences on (text-based) content pages, we filter out fixations on search engine result pages and video platforms.
Before (\emph{Pre}) and after (\emph{Post}) the web search, (1) a multiple choice test and (2) a written essay have been performed, further denoted as \emph{MCQ} and \emph{Essay}, respectively.
For \emph{MCQ}, the correct answers have been counted; for \emph{Essay}, the authors counted the correctly identified concepts, allowing us to define the knowledge gain as $\emph{KG}{=}\emph{Post} {-} \emph{Pre}$.

\paragraph{Metric:}
We divide the participants into two groups per setting: a \emph{Low} group if they scored lower than the average \emph{KG} across participants or \emph{High} if higher.
The distributions for \emph{MCQ} and \emph{Essay} for \emph{Pre}, \emph{Post}, and \emph{KG} are as follows:
\begin{itemize}
    \item \emph{Pre}: \num{63} participants are in \emph{Low} for \emph{MCQ}, \num{57} in \emph{Essay}, and \num{42} assigned \emph{Low} for both; respectively \num{43}, \num{49} and \num{42} for \emph{High}
    \item \emph{Post}: \num{54} participants are in \emph{Low} for \emph{MCQ}, \num{54} in \emph{Essay}, and \num{30} assigned \emph{Low} for both; respectively \num{52}, \num{52} and \num{28} for \emph{High}
    \item \emph{KG}: \num{66} participants are in \emph{Low} for \emph{MCQ}, \num{46} in \emph{Essay}, and \num{32} assigned \emph{Low} for both; respectively \num{40}, \num{60} and \num{26} for \emph{High} 
\end{itemize}
Since the overlaps between \emph{MCQ} and \emph{Essay} are not very strong, we report the results for both test forms.
As the feature values are not normally distributed, we use the parameter-free Mann-Whitney U test to determine significance.

\paragraph{Results:}
\input{tables/0_combined_tables}
To answer the first research question (1), we analyze the variables regarding the total reading duration on content pages, the average
reading duration
per content page and finally the number of reading fixations per content page (see first part of Table~\ref{tab:1_comparism_nilavra}).
We observe that learners from the \emph{Low} group read for significantly shorter durations and had fewer fixations per content page (for \emph{Essay}) compared to the \emph{High} group.
The observation that the total reading time is increased for learners with high knowledge gain based on \emph{Essay} is consistent with Bhattacharya and Gwizdka~\cite{Bhattacharya2019}.
However, our results do not reveal that learners with low knowledge gain have read longer in total or on average, nor do they have more reading fixations.
This suggests that these observations may not be generalizable and require further research.

For the second research question (2), we consider several extracted features for the pretests.
The results show that fewer features differ significantly in prior knowledge compared to \emph{Post} and \emph{KG}.
However, significant differences in average and maximum y-coordinate of reading fixations exist for both \emph{MCQ} and \emph{Essay} (see Table~\ref{tab:1_comparism_nilavra} after double line).
As there is no significant difference in the number of visited content pages, this suggests that readers with more prior knowledge engage more deeply with web pages than those with less prior knowledge.
This behavior does not seem to imply a significant difference in \emph{KG}.

To answer the last research question (3), we analyze a subset of eye-tracking features\footnote{Complete lists of results: \url{https://github.com/TIBHannover/reading_web_search/tree/master/results}} regarding \emph{KG} (see second part of Table~\ref{tab:1_comparism_nilavra}).
Our results indicate that more successful learners read longer in total on content pages (in terms of the number of fixations and reading sequences).
Furthermore, learners that processed more words in general processed more unique words, achieving a noticeably better result in the \emph{Essay} scores.
Furthermore, a difference in the time learners read until regressions occur can be seen for both scores.
Combined with the higher reading speed
for successful learners (for \emph{MCQ}), this may indicate that it can be more efficient for learning to skim web pages for new information and regress as required, as opposed to general slow and thorough reading.

\section{Conclusions}\label{sec:conclusions}
\paragraph{Summary:}
In this paper, we have presented the extension of an existing reading model~\cite{Cole2011,Cole2013,Bhattacharya2019}. 
While prior work solely relied on fixations, we incorporated additional information on the positions of text on web pages and sequences across line breaks.
To determine the impact on learning outcomes, we calculated eye-tracking features for textual content pages, such as the total number of read words and the read words per second.
We evaluated the impact of the features on pretest, posttest, and knowledge gain scores from multiple choice and essay assessments.
Our findings revealed that learners with higher knowledge gain had spent more time reading and had an increased number of fixations, contradicting the opposite observation of Bhattacharya and Gwizdka~\cite{Bhattacharya2019}.
We also found that learners who had higher prior knowledge, read more deeply on content pages, without necessarily experiencing higher knowledge gains.
Finally, we observed that learners with high knowledge gains read more intensively in terms of the number of words read and unique words.
In addition, reading speed and time to backward regressions were also increased for more successful learners.

\paragraph{Limitations:}
Currently, our model only works for languages with a left-to-right reading direction, but an adaptation is possible.
Furthermore, while we calculated the influence of the computed features on knowledge gain, we did not evaluate the reading model itself,
i.e., how accurately reading sequences are recognized by our model.
System parameters, such as the minimum number of words for reading sequences, were chosen based on commonsense reasoning.
In the future, empirical evidence should be provided to support these choices.
Finally, the scope of this study was limited to the textual content of web pages.
As prior work underlined, learning processes are often supported by other visual formats, such as images and videos~\cite{Otto2021,Pardi2023}.
Their impact on gaze direction and knowledge gain provide interesting directions for future work.

\bibliographystyle{splncs04}
\bibliography{bibliography}

\appendix

\end{document}

%% file: tables/0_combined_tables.tex
\begin{table}[!b]
\setlength{\tabcolsep}{1.2pt}
\setlength\extrarowheight{0pt}
\centering
\fontsize{7}{10}\selectfont
\def\arraystretch{0.8}
\caption{Mean and p-value based on a Mann Whitney U test in multiple-choice test (MCQ) and essay for knowledge gain and prior knowledge (after double line), respectively. %Approaching 
Differences with $0.05 \leq p < 0.1$ are \underline{underlined}, significant differences with $0.01 < p \leq 0.05$ \textbf{bold}, and very significant differences with $p < 0.01$ \underline{\textbf{both}}.}
\label{tab:1_comparism_nilavra}
\begin{tabular}{lccccccp{.465\linewidth}}
\toprule
 & \multicolumn{3}{c}{\textbf{MCQ}} & \multicolumn{3}{c}{\textbf{Essay}} & \textbf{} \\
\textbf{Feature} & \textbf{Low} & \textbf{High} & \textbf{p} & \textbf{Low} & \textbf{High} & \textbf{p} & \textbf{Description} \\
\midrule
\textit{sum\_RFix\_dur} & 186.9& 177.0 & .747 & 162.1& 199.3& \textbf{.040} & \textit{Sum of reading fixation durations (in ms) on content pages} \\
\textit{avg\_RFix\_dur} & 40.90 & 34.83 & .722 & 33.00 & 42.91 & .130 & \textit{Sum of reading fixation durations (in ms), averaged across content pages} \\
\textit{avg\_n\_RFix} & 84.58 & 79.44 & .582 & 65.35 & 95.89& \textbf{.045} & \textit{Count of reading fixations, averaged across content pages} \\
\midrule
\textit{n\_CP\_visited} & 6.59 & 6.42 & .768 & 6.15 & 6.82 & .170 & \textit{number of visited content pages} \\
\textit{avg\_Fix\_dur} & 471.1 & 432.9 & \textbf{.015} & 476.3 & 441.7 & \underline{.059} & \textit{average duration of any fixations (in ms)} \\
\textit{dur\_per\_RSeq} & 1941& 1774& \underline{.079} & 1969& 1808& \textbf{.046} & \textit{average duration of reading sequences (in ms)} \\
\textit{n\_RSeq} & 99.05 & 101.4 & .611 & 82.83 & 113.1 & \underline{\textbf{.006}} & \textit{number of reading sequences} \\
\textit{avg\_RFix\_dur} & 489.8& 447.3 & \textbf{.024} & 497.0& 455.9 & \textbf{.046} & \textit{average duration of reading fixations (in ms)} \\
\textit{n\_RFix} & 392.4& 406.0& .560 & 326.9& 451.7& \underline{\textbf{.009}} & \textit{number of reading fixations} \\
\textit{n\_Reg} & 25.76 & 26.32 & .351 & 20.48 & 30.18 & \textbf{.021} & \textit{number of backward regressions} \\
\textit{n\_Reg\_per\_sec} & 0.13 & 0.15 & \textbf{.026} & 0.13 & 0.14 & \underline{.099} & \textit{number of regressions per second} \\
\textit{n\_unique\_word} & 528.1& 533.4& .553 & 454.7& 588.0& \textbf{.014} & \textit{total number of unique words read} \\
\textit{n\_words} & 692.6& 727.6& .510 & 587.3& 796.7& \textbf{.010} & \textit{total count of words read (including duplicates, e.g., by regressions)} \\
\textit{words\_per\_sec} & 3.82 & 4.12 & \textbf{.023} & 3.84 & 4.00 & .151 & \textit{read words per second} \\
\midrule
\midrule
\textit{max\_y\_of\_RFix} & 2889& 3644& \textbf{.016} & 2790& 3667& \underline{.057} & \textit{maximum y-position of reading fixations} \\
\textit{avg\_y\_of\_RFix} & 1160& 1272& \underline{.073} & 1123& 1301& \underline{\textbf{.005}} & \textit{average y-position of reading fixations} \\
\bottomrule
\end{tabular}
\end{table}